\documentclass[12pt,A4]{article}
\usepackage{amsmath,amsfonts}
\usepackage[numbers,sort&compress]{natbib}
\usepackage[top=3.2cm,bottom=3.5cm,inner=2cm,outer=2cm]{geometry}

\begin{document}

\title{Nonstandard Fourier Pseudospectral Time Domain \\ (PSTD) Schemes for Partial Differential Equations}

\author{Bradley E. Treeby, Elliott S. Wise, B. T. Cox \\ \\ Department of Medical Physics and Biomedical Engineering \\ University College London, United Kingdom \\ {\tt b.treeby@ucl.ac.uk}}

\maketitle

\abstract{A class of nonstandard pseudospectral time domain (PSTD) schemes for solving time-dependent hyperbolic and parabolic partial differential equations (PDEs) is introduced. These schemes use the Fourier collocation spectral method to compute spatial gradients and a nonstandard finite difference scheme to integrate forwards in time. The modified denominator function that makes the finite difference time scheme exact is transformed into the spatial frequency domain or $k$-space using the dispersion relation for the governing PDE. This allows the correction factor to be applied in the spatial frequency domain as part of the spatial gradient calculation. The derived schemes can be formulated to be unconditionally stable, and apply to PDEs in any space dimension. Examples of the resulting nonstandard PSTD schemes for several PDEs are given, including the wave equation, diffusion equation, and convection-diffusion equation.}


\section{Introduction}

Over the last decade, nonstandard finite difference methods have been used for the numerical solution of a wide range of differential equations \cite{Mickens1994,Patidar2005}. These schemes are constructed by modifying the denominator of conventional finite difference formulae such that the phase error introduced by the discretisation is eliminated \cite{Mickens2006}. This is particularly useful in the context of large-scale problems, as it alleviates the need to use dense spatial and temporal grids to avoid the accumulation of numerical dispersion. However, while the idea behind nonstandard finite difference methods is very appealing, many of the existing schemes have only been derived in one space dimension, and are not easily generalisable to higher dimensions \cite{Mickens2002a}. Moreover, for solving time dependent partial differential equations (PDEs), the denominator functions typically depend on the Fourier representation of the independent variables (e.g., the spatial or temporal frequency). To evaluate the resulting numerical scheme in the time and space domain, a single value for these variables must be chosen, which means the schemes are only exact for the chosen Fourier component \cite{Cole2000}.

Recently, modified pseudospectral time domain (PSTD) schemes for solving time dependent PDEs in acoustics and electromagnetics have also been proposed \cite{Haber1973,Mast2001,Treeby2012,Vay2013}. These schemes use the Fourier pseudospectral (or collocation) method to compute spatial derivatives, and a corrected finite difference scheme to integrate forwards in time. In contrast to nonstandard finite difference schemes, the correction to the finite difference time scheme is applied in the {\em spatial} frequency domain (often referred to as $k$-space) as part of the spectral calculation of spatial derivatives. The appropriate correction factor is derived by considering the Green's function solution for the governing PDE \cite{Mast2001,Cox2005}, and eliminates the phase error introduced by the finite difference time step.  For linear problems, these methods have the advantage of being exact for all spatial and temporal frequencies up to the Nyquist limit, and can be applied in any space dimension.

Here, the idea of nonstandard PSTD schemes is generalised to a broader class of hyperbolic and parabolic PDEs, and the derivation of these schemes from nonstandard finite difference methods is shown. The schemes are based on the Fourier pseudospectral method for discretising spatial gradients, and a nonstandard finite difference scheme for time integration. The modified denominator function for the nonstandard finite difference time scheme is transformed into the spatial frequency domain (or $k$-space) using the dispersion relation for the governing PDE. This allows the correction term to be applied in the spatial frequency domain as part of the spatial gradient calculation. The general formulation of the nonstandard PSTD schemes for constant coefficient PDEs is given in Sec.\ 2, with several examples given in Sec.\ 3. Application to the case of non-constant coefficient PDEs is discussed in Sec.\ 4. Discussion and summary are then provided in Sec.\ 5.


\section{Formulation of nonstandard PSTD schemes for constant-coefficient PDEs}

\subsection{Problem formulation}

We are interested in deriving exact explicit schemes for numerically solving inhomogeneous linear hyperbolic and parabolic PDEs in an unbounded domain, where the PDEs are in the form
\begin{equation}
\frac{\partial^a}{\partial t^a}  u(\mathbf{x}, t)= \mathrm{L} u(\mathbf{x}, t) + S(\mathbf{x}, t) \enspace.
\label{eq_general_pde}
\end{equation}
Here the PDE has a single temporal derivative of positive integer order $a \in \left\{ 1, 2, 3, 4 \right\}$, $u$ is a scalar field variable defined as a function of position in Euclidean space $\mathbf{x} \in \mathbb{R}^d,\;d \in \left\{ 1, 2, 3 \right\} $ and time $t \in \mathbb{R}$, $\mathrm{L}$ is a constant-coefficient spatial differential operator (the non-constant coefficient case is discussed in Sec.\ 4), and $S(\mathbf{x}, t)$ is a source term. There are many PDEs of practical interest that are in this form, including the linear transport equation, wave equation, and diffusion equation.


\subsection{Exact finite difference time schemes}

Consider the solution $u(\mathbf{x}, t)$ of the continuous PDE. Using the following definition for the non-unitary Fourier transform, this can be expanded in a Fourier basis as
\begin{align}
u(\mathbf{x}, t) &= \int_{-\infty}^{\infty} \tilde{u}(\mathbf{x}, \omega) e^{-i\omega t} d\omega \enspace, \nonumber\\
\tilde{u}(\mathbf{x}, \omega) &= \frac{1}{2 \pi}\int_{-\infty}^{\infty} u(\mathbf{x}, t) e^{i\omega t} dt \enspace,
\end{align}
where $\omega \in \mathbb{R}$, and the tilde $\tilde{u}$ is used to indicate a variable in the temporal frequency domain. Using this expansion, the temporal derivative of $u(\mathbf{x}, t)$ can be written exactly in Fourier space as
\begin{equation}
\frac{\partial^a}{\partial t^a} u(\mathbf{x}, t) = \int_{-\infty}^{\infty} \tilde{u}(\mathbf{x}, \omega) (-i\omega)^a e^{-i\omega t} d\omega \enspace.
\label{eq_exact_derivative_soln}
\end{equation}
This result can be used to derive nonstandard finite difference schemes for different derivative orders. For a first-order time derivative, where $a = 1$, using an explicit first-order accurate forward difference (i.e., the forward Euler method), the temporal derivative can be approximated as
\begin{equation}
\frac{\partial}{\partial t} u(\mathbf{x}, t) \approx \frac{u(\mathbf{x}, t + \Delta t) - u(\mathbf{x}, t)}{\Delta t} = \int_{-\infty}^{\infty} \tilde{u}(\mathbf{x}, \omega) \left( \frac{e^{-i\omega \Delta t} - 1}{\Delta t} \right)  e^{-i\omega t} d\omega   \enspace.
\label{eq_first_order_fd_soln}
\end{equation}
Comparing Eqs.\ \eqref{eq_exact_derivative_soln} and \eqref{eq_first_order_fd_soln}, the approximate solution for the derivative calculated by the finite difference scheme differs from the exact solution by a factor of
\begin{equation}
\kappa_t \left( \omega \right)= \frac{1 - e^{-i\omega \Delta t}}{i \omega \Delta t} \enspace.
\label{eq_general_phase_correction_dt}
\end{equation}
This difference introduces unwanted numerical dispersion or phase error into the derivative calculation. Practically, this can be controlled by changing the size of the time step $\Delta t$ (where $\kappa_t \rightarrow 1$ as $\Delta t \rightarrow 0$). However, as the error in the finite difference approximation is known exactly in closed form, it can be introduced into the finite difference scheme as a correction factor. This approach is referred to as a {\em nonstandard finite difference scheme}, and is discussed in detail by Mickens and others \cite{Mickens2002a,Patidar2005,Mickens2006}. For a first-order time derivative, the resulting nonstandard finite difference scheme is given by
\begin{equation}
\frac{\partial}{\partial t} u(\mathbf{x}, t) = \frac{u(\mathbf{x}, t + \Delta t) - u(\mathbf{x}, t)}{\kappa_t \left( \omega \right) \Delta t} \enspace,\end{equation}
where $\kappa_t (\omega)\Delta t$ is sometimes referred to as the denominator function \cite{Mickens2006}.

The same approach can be used for higher-order time derivatives. For example, for a second-order time derivative, where $a = 2$, using an explicit second-order accurate central difference, the temporal derivative can be approximated as
\begin{align}
\frac{\partial^2}{\partial t^2} u(\mathbf{x}, t) &\approx \frac{u(\mathbf{x}, t + \Delta t) - 2 u(\mathbf{x}, t) + u(\mathbf{x}, t - \Delta t)}{\Delta t^2} \nonumber\\
&=  \int_{-\infty}^{\infty} \tilde{u}(\mathbf{x}, \omega) \left( \frac{e^{-i\omega \Delta t} - 2 + e^{i\omega \Delta t}}{\Delta t^2} \right) e^{-i\omega t} d\omega \enspace.
\end{align}
Compared to the exact solution given in Eq.\ \eqref{eq_exact_derivative_soln}, the approximate solution to the derivative calculated by the finite difference scheme differs from the exact solution by a factor of
\begin{equation}
\kappa_{tt} \left( \omega \right) = \frac{e^{-i\omega \Delta t} - 2 + e^{i\omega \Delta t}}{-\omega^2 \Delta t^2} = \mathrm{sinc}^2(\omega \Delta t/2)\enspace,
\label{eq_general_phase_correction_dtt}
\end{equation}
where $\mathrm{sinc}(x)= \sin(x)/x$. Again, this factor can be introduced as a correction to the finite difference time scheme to give an exact time stepping scheme
\begin{equation}
\frac{\partial^2}{\partial t^2} u(\mathbf{x}, t) = \frac{u(\mathbf{x}, t + \Delta t) - 2 u(\mathbf{x}, t) + u(\mathbf{x}, t - \Delta t)}{\kappa_{tt} \left( \omega \right) \Delta t^2} \enspace.
\end{equation}
Similar schemes can also be derived for higher order derivatives. Note, following Mickens' rules for constructing stable nonstandard finite difference schemes, the derivatives must be discretised using a finite difference scheme of the same order accuracy as the order of derivative \cite{Mickens2002a,Patidar2005}. 


\subsection{Exact pseudospectral time domain schemes}

For PDEs in the form of Eq.\ \eqref{eq_general_pde}, an exact time-stepping solution can be formed by combining the nonstandard finite difference time schemes discussed in the previous section with the spectral calculation of the spatial derivatives using a Fourier basis. The $d$-dimensional spatial Fourier transform is defined as
\begin{align}
u(\mathbf{x}, t) &= \int_{\mathbb{R}^d} \hat{u}(\mathbf{k}, t) e^{i \mathbf{k} \cdot \mathbf{x}} d \mathbf{k} \equiv \mathcal{F}^{-1} \left\{ \hat{u}(\mathbf{k}, t) \right\} \enspace, \nonumber\\
\hat{u}(\mathbf{k}, t) &= \frac{1}{(2 \pi)^d}\int_{\mathbb{R}^d} u(\mathbf{x}, t) e^{- i \mathbf{k} \cdot \mathbf{x}} d\mathbf{x} \equiv \mathcal{F} \left\{ u(\mathbf{x}, t)  \right\}  \enspace,
\end{align}
where $\mathbf{k} \in \mathbb{R}^d$ is the wave vector, and the hat symbol $\hat{u}$ is used to indicate a variable in the spatial frequency domain. The Fourier representation of $\mathrm{L} u(\mathbf{x}, t)$ can then be written as
\begin{equation}
\mathrm{L} u(\mathbf{x}, t) = \int_{\mathbb{R}^d}  \lambda(\mathbf{k}) \hat{u}(\mathbf{k}, t) e^{i \mathbf{k} \cdot \mathbf{x}} d \mathbf{k} = \mathcal{F}^{-1} \left\{ \lambda(\mathbf{k}) \mathcal{F} \left\{ u(\mathbf{x}, t)  \right\}  \right\}  \enspace,
\label{eq_spectral_derivative}
\end{equation}
where $\lambda(\mathbf{k})$ is the Fourier representation of the differential operator $\mathrm{L}$. For example, if $\mathrm{L} = \nabla^2$ then $\lambda(\mathbf{k}) = i \mathbf{k} \cdot i\mathbf{k} \equiv -k^2$. Combining Eq.\ \eqref{eq_spectral_derivative} with the nonstandard finite difference time scheme from Eq.\ \eqref{eq_general_phase_correction_dt} then gives the following exact time-stepping solution to Eq.\ \eqref{eq_general_pde} for $a=1$
 \begin{equation}
\frac{u(\mathbf{x}, t+\Delta t) - u(\mathbf{x}, t)}{\kappa_t\left( \omega \right) \Delta t} = \mathcal{F}^{-1} \left\{ \lambda(\mathbf{k}) \mathcal{F} \left\{ u(\mathbf{x}, t) \right\} \right\} + S(\mathbf{x}, t) \enspace.
\label{eq_exact_soln_a1}
\end{equation}
Similarly, combining Eq.\ \eqref{eq_spectral_derivative} with Eq.\ \eqref{eq_general_phase_correction_dtt} gives an exact time-stepping solution to Eq.\ \eqref{eq_general_pde} for $a=2$
\begin{equation}
\frac{u(\mathbf{x}, t + \Delta t) - 2 u(\mathbf{x}, t) + u(\mathbf{x}, t - \Delta t)}{\kappa_{tt} \left( \omega \right) \Delta t^2} = \mathcal{F}^{-1} \left\{ \lambda(\mathbf{k}) \mathcal{F} \left\{ u(\mathbf{x}, t) \right\} \right\}  + S(\mathbf{x}, t)\enspace.
\label{eq_exact_soln_a2}
\end{equation}

Analogous to other nonstandard finite difference methods, the primary limitation of solutions to Eq.\ \eqref{eq_general_pde} in the form of Eqs.\ \eqref{eq_exact_soln_a1} and \eqref{eq_exact_soln_a2} is that the denominator function used in the nonstandard finite difference time scheme depends on the temporal frequency $\omega$. As the solution is expressed in the time-domain, a single value must be chosen for $\omega$, meaning the solution is only exact for a single Fourier component. The approach proposed here is to use the dispersion relation between the temporal frequency $\omega$ and the spatial frequency $\mathbf{k}$ (obtained by taking the temporal and spatial Fourier transform of the homogeneous governing PDE) and substitute this into the denominator function. This gives a correction term that depends on $\mathbf{k}$ rather than $\omega$, which can then be applied in the spatial frequency domain as part of the spectral spatial gradient calculation. 

Starting with the case for $a = 1$, the dispersion relation for Eq.\ \eqref{eq_general_pde} can be written in general form as
\begin{equation}
\omega = i \lambda(\mathbf{k}) \enspace.
\label{eq_1st_order_general_dispersion_relation}
\end{equation}
Substituting this into Eq.\ \eqref{eq_general_phase_correction_dt} then gives a correction term expressed in the spatial frequency domain or $k$-space
\begin{equation}
\hat{\kappa}_t \left( \mathbf{k} \right) = \frac{e^{\lambda(\mathbf{k}) \Delta t} - 1}{\lambda(\mathbf{k}) \Delta t} \enspace.
\label{eq_kappa_t_kspace}
\end{equation}
Here the hat $\hat{\kappa}_t$ is used to indicate that the correction term is expressed as a function of $\mathbf{k}$ rather than $\omega$. Using Eq.\ \eqref{eq_exact_soln_a1}, an exact time-stepping solution to Eq.\ \eqref{eq_general_pde} can then be written as
\begin{equation}
\frac{u(\mathbf{x}, t+\Delta t) - u(\mathbf{x}, t)}{\Delta t} = \mathcal{F}^{-1} \left\{ \hat{\kappa}_t \left( \mathbf{k} \right) \left(  \lambda(\mathbf{k})  \mathcal{F} \left\{ u(\mathbf{x}, t) \right\} +  \mathcal{F} \left\{ S(\mathbf{x}, t) \right\} \right) \right\} \enspace.
\label{eq_first_order_time_deriv_xt}
\end{equation}
Unlike Eq.\ \eqref{eq_exact_soln_a1}, this doesn't depend on the temporal frequency $\omega$, and is exact in $\mathbb{R}^d$ for all $\omega, \mathbf{k}$. Thus, for any PDE in the form of Eq.\ \eqref{eq_general_pde} with $a=1$, an exact time stepping solution can be obtained by using the Fourier representation $\lambda(\mathbf{k})$ of the spatial differential operator $\mathrm{L}$ and substituting this into Eqs.\ \eqref{eq_kappa_t_kspace} and \eqref{eq_first_order_time_deriv_xt}. 

For $a = 2$,  the dispersion relation for Eq.\ \eqref{eq_general_pde} can be written in general form as
\begin{equation}
\omega = \pm \sqrt{ - \lambda(\mathbf{k})} \enspace.
\label{eq_2nd_order_general_dispersion_relation}
\end{equation}
Substituting this into Eq.\ \eqref{eq_general_phase_correction_dtt}, then gives the $k$-space correction term as
\begin{equation}
\hat{\kappa}_{tt} \left( \mathbf{k} \right) = \mathrm{sinc}^2 \left( \sqrt{ - \lambda(\mathbf{k})} \Delta t/2 \right) \enspace.
\label{eq_kappa_tt_kspace}
\end{equation}
Using Eq.\ \eqref{eq_exact_soln_a2}, an exact time-stepping solution to Eq.\ \eqref{eq_general_pde} can then be written as
\begin{equation}
\frac{u(\mathbf{x}, t + \Delta t) - 2 u(\mathbf{x}, t) + u(\mathbf{x}, t - \Delta t)}{\Delta t^2} = \mathcal{F}^{-1} \left\{ \hat{\kappa}_{tt}\left( \mathbf{k} \right) \left(  \lambda(\mathbf{k}) \mathcal{F} \left\{ u(\mathbf{x}, t) \right\}  + \mathcal{F} \left\{ S(\mathbf{x}, t) \right\} \right) \right\}\enspace.
\label{eq_second_order_scheme_xt}
\end{equation}
Thus, for any PDE in the form of Eq.\ \eqref{eq_general_pde} with $a=2$, an exact time stepping solution can be obtained by using the Fourier representation $\lambda(\mathbf{k})$ of the spatial differential operator $\mathrm{L}$ and substituting this into Eqs.\ \eqref{eq_kappa_tt_kspace} and \eqref{eq_second_order_scheme_xt}. Similar expressions can be derived for $a=3$ and $a=4$ (the highest order time derivative for which an explicit dispersion relation can always be formed). As these schemes are derived from nonstandard finite difference time schemes and the Fourier pseudospectral method, they are refered to herein as {\em nonstandard PSTD schemes}. 


\subsection{Stability}
The nonstandard PSTD schemes given in Eqs.\ \eqref{eq_first_order_time_deriv_xt} and \eqref{eq_second_order_scheme_xt} can be shown to be unconditionally stable. Beginning with the $a=1$ case, Eq.\ \eqref{eq_first_order_time_deriv_xt} can be written in the spatial frequency domain as 
\begin{equation}
\frac{\hat{u}^{n+1} - \hat{u}^n}{\Delta t} = \hat{\kappa}_t \left( \mathbf{k} \right) \lambda(\mathbf{k}) \hat{u}^n \enspace,
\label{eq_first_order_time_deriv}
\end{equation}
where $\hat{u}^n \equiv \mathcal{F} \left\{ u(\mathbf{x}, t ) \right\}$ and $\hat{u}^{n+1} \equiv \mathcal{F} \left\{ u(\mathbf{x}, t + \Delta t ) \right\}$. Rearranging Eq.\ \eqref{eq_first_order_time_deriv} gives an update equation of the form
\begin{equation}
\hat{u}^{n+1} = \left( 1 +  \Delta t \lambda(\mathbf{k}) \hat{\kappa}_t \left( \mathbf{k} \right)  \right)\hat{u}^n \enspace.
\end{equation}
Substituting for the $k$-space correction from Eq.\ \eqref{eq_kappa_t_kspace} then gives
\begin{equation}
\hat{u}^{n+1} =  e^{\lambda(\mathbf{k}) \Delta t} \hat{u}^n \enspace.
\end{equation}
This scheme is stable when $|e^{\lambda(\mathbf{k}) \Delta t}| \le 1$. This is always true provided that the real part of $\lambda(\mathbf{k})$ is negative or zero for all $\mathbf{k}$. 

For $a = 2$, Eq.\ \eqref{eq_second_order_scheme_xt} can be written in the spatial frequency domain as
\begin{equation}
\frac{\hat{u}^{n+1} - 2 \hat{u}^n + \hat{u}^{n-1}}{\Delta t^2} =   \lambda(\mathbf{k}) \hat{\kappa}_{tt} \left( \mathbf{k} \right) \hat{u}^n \enspace.
\end{equation}
Substituting for the $k$-space correction term from Eq.\ \eqref{eq_kappa_tt_kspace} then leads to
\begin{equation}
\hat{u}^{n+1} - 2 \hat{u}^n + \hat{u}^{n-1}= - b^2 \hat{u}^n \enspace,
\label{eq_second_order_scheme_stability}
\end{equation}
where $b =  2 \sin ( \sqrt{-\lambda(\mathbf{k})} \Delta t/2) $. The range of values of $b$ for which this scheme generates a stable sequence $\ldots,\hat{u}^{n-1}, \hat{u}^n, \hat{u}^{n+1},\ldots$ can be found by assuming the solution at timestep $n$ has the form $\hat{u}^n = (A)^n B$. Substituting this into Eq.\ \eqref{eq_second_order_scheme_stability} leads to the characteristic quadratic equation
\begin{equation}
\label{EQ_characteristic_equation}
A^2 + (b^2-2)A + 1 = 0 \enspace,
\end{equation}
for which the two solutions are
\begin{equation}
A_{1,2} = \frac{-(b^2-2)\pm \sqrt{(b^2-2)^2 - 4}}{2} \enspace.
\end{equation}
This scheme is stable when $|A|\le 1$, which occurs when $|b|\le 2$. This is always true provided that $\lambda(\mathbf{k})$ is a negative real number or zero for all $\mathbf{k}$.


\section{\label{sec_examples}Examples of nonstandard PSTD schemes}

The nonstandard PSTD schemes derived in the previous section can be used to generate exact and unconditionally stable solutions to any hyperbolic or parabolic PDE in the form of Eq.\ \eqref{eq_general_pde}. Several examples are given in the following sections.


\subsection{Wave equation}

The linearised constant-coefficient wave equation is given by
\begin{equation}
\frac{\partial^2 u(\mathbf{x}, t)}{\partial t^2} = c_0^2 \nabla^2 u(\mathbf{x}, t) \enspace,
\end{equation}
where $c_0 \in \mathbb{R}_{>0}$. The Fourier representation of the spatial differential operator is $\lambda(\mathbf{k}) = -c_0^2 k^2$. From 
Eq.\ \eqref{eq_kappa_tt_kspace} the $k$-space correction term is
\begin{equation}
\hat{\kappa}_{tt}(\mathbf{k}) =  \mathrm{sinc}^2(c_0 k \Delta t/2)\enspace,
\label{eq_kspace_correction_wave_equation}
\end{equation}
so the nonstandard  PSTD scheme becomes
\begin{equation}
\frac{u^{n+1} - 2 u^n + u^{n-1}}{\Delta t^2} = \mathcal{F}^{-1} \{ - \hat{\kappa}_{tt}(\mathbf{k})  c_0^2 k^2 \, \mathcal{F}\{ u^n \}   \} \enspace.
\label{eq_exact_PSTD_wave_equation}
\end{equation}
As $c_0^2 \in \mathbb{R}_{>0}$ and $k^2 \in \mathbb{R}_{\ge0}$, this means $\lambda(\mathbf{k}) \in \mathbb{R}_{\le 0} \; \forall \, \mathbf{k}$ and the scheme is unconditionally stable. This particular scheme was first introduced by Bojarski \cite{Bojarski1985} (although derived using a different approach and given in a different form), and has since been used by a number of other authors, e.g., \cite{Mast2001,Cox2005}. A similar scheme was also discussed earlier by Fornberg \& Whitham in relation to the high-$k$ approximation of the Korteweg-de Vries equation \cite{Fornberg1978}, and Haber {\em et al.,} in relation to electromagnetic waves \cite{Haber1973}. 


\subsection{Dispersive wave equation}

A constant-coefficient  dispersive wave equation is given by \cite{Treeby2010a}
\begin{equation}
\frac{\partial^2 u(\mathbf{x}, t)}{\partial t^2} = c_0^2 \nabla^2 u(\mathbf{x}, t) + \eta (-\nabla^2)^{(y+1)/2} u(\mathbf{x}, t)\enspace,
\end{equation}
where $(-\nabla^2)^a$ is the fractional Laplacian, $c_0, \eta \in \mathbb{R}_{>0}$, $y \in [1,2]$, and generally $\eta \ll c_0$. The Fourier representation of the spatial differential operator is $\lambda(\mathbf{k}) = -c_0^2 k^2 + \eta k^{y+1}$. From 
Eq.\ \eqref{eq_kappa_tt_kspace} the $k$-space correction term is
\begin{equation}
\hat{\kappa}_{tt} (\mathbf{k}) =  \mathrm{sinc}^2\left(\sqrt{ c_0^2 k^2 - \eta k^{y+1} } \Delta t/2\right)\enspace,
\end{equation}
so the nonstandard PSTD scheme becomes
\begin{equation}
\frac{u^{n+1} - 2 u^n + u^{n-1}}{\Delta t^2} = \mathcal{F}^{-1} \{ - \hat{\kappa}_{tt}(\mathbf{k}) \left( c_0^2 k^2 - \eta k^{y+1} \right) \mathcal{F}\{ u^n \}   \} \enspace.
\label{eq_exact_PSTD_wave_equation}
\end{equation}
When $\eta < c_0^2 k^{1-y} \, \forall \, \mathbf{k} $, then $\lambda(\mathbf{k}) \in \mathbb{R}_{\le 0} \; \forall \, \mathbf{k}$, and the scheme is unconditionally stable. 


\subsection{Diffusion equation}

The constant-coefficient diffusion equation is given by
\begin{equation}
\frac{\partial u(\mathbf{x}, t)}{\partial t} = D \nabla^2 u(\mathbf{x}, t) \enspace,
\end{equation}
where $D \in \mathbb{R}_{>0}$. The Fourier representation of the spatial differential operator is $\lambda(\mathbf{k}) = -D k^2$. From Eq.\ \eqref{eq_kappa_t_kspace} the $k$-space correction term is
\begin{equation}
\hat{\kappa}_t(\mathbf{k}) = \frac{1 - e^{- D k^2 \Delta t}}{ D k^2 \Delta t} \enspace,
\end{equation}
so the nonstandard PSTD scheme becomes
\begin{equation}
\frac{u^{n+1} - u^{n}}{\Delta t} = \mathcal{F}^{-1} \left\{ - \hat{\kappa}_t (\mathbf{k})D k^2 \, \mathcal{F}\left\{ u^n \right\} \right\} \enspace.
\end{equation}
As $\lambda(\mathbf{k}) \in \mathbb{R}_{\le 0}  \; \forall \, \mathbf{k}$, the scheme is unconditionally stable.


\subsection{Bioheat / optical diffusion equation}

The constant-coefficient optical diffusion equation or bioheat equation (these are equivalent) is given by
\begin{equation}
\frac{\partial u(\mathbf{x}, t)}{\partial t} = D \nabla^2 u(\mathbf{x}, t) - P u(\mathbf{x}, t) + S(\mathbf{x}, t)\enspace,
\end{equation}
where $D, P \in \mathbb{R}_{>0}$. The Fourier representation of the spatial differential operator is $\lambda(\mathbf{k}) = -D k^2 - P$. From Eq.\ \eqref{eq_kappa_t_kspace} the $k$-space correction term is
\begin{equation}
\hat{\kappa}_t(\mathbf{k}) = \frac{1 - e^{- \left( D k^2  + P \right) \Delta t}}{ \left( D k^2  + P \right) \Delta t} \enspace,
\end{equation}
so the nonstandard PSTD scheme becomes
\begin{equation}
\frac{u^{n+1} - u^{n}}{\Delta t} = \mathcal{F}^{-1} \left\{ \hat{\kappa}_t (\mathbf{k}) \left[  \left( - Dk^2 - P \right) \mathcal{F}\left\{ u^n \right\} + \mathcal{F}\left\{ S^n \right\}  \right] \right\} \enspace.
\end{equation}
As $\lambda(\mathbf{k}) \in \mathbb{R}_{\le 0}  \; \forall \, \mathbf{k}$, the scheme is unconditionally stable. A similar scheme was proposed in \cite{Gao1995}.


\subsection{Convection-diffusion equation}

The convection-diffusion equation is given by
\begin{equation}
\frac{\partial u(\mathbf{x}, t)}{\partial t} = D \nabla^2 u(\mathbf{x}, t) - \mathbf{C} \cdot \nabla u(\mathbf{x}, t) \enspace,
\end{equation}
where $D \in \mathbb{R}_{>0}$ and $\mathbf{C} \in \mathbb{R}^d$. The Fourier representation of the spatial differential operator is $\lambda(\mathbf{k}) = -D k^2  - i \mathbf{C} \cdot \mathbf{k}$. From Eq.\ \eqref{eq_kappa_t_kspace} the $k$-space correction term is
\begin{equation}
\hat{\kappa}_t (\mathbf{k})= \frac{1 - e^{-Dk^2 \Delta t -  i (\mathbf{C} \cdot \mathbf{k} )\Delta t}}{Dk^2 \Delta t +  i (\mathbf{C} \cdot \mathbf{k} )\Delta t } \enspace,
\end{equation}
so the nonstandard PSTD scheme becomes
\begin{equation}
\frac{u^{n+1} - u^{n}}{\Delta t} = \mathcal{F}^{-1} \left\{ - \hat{\kappa}_t (\mathbf{k})\left( D k^2  + i \mathbf{C} \cdot \mathbf{k} \right) \mathcal{F} \left\{ u^n \right\} \right\} \enspace.
\end{equation}
As $\mathrm{Re}(\lambda(\mathbf{k})) \in \mathbb{R}_{\le0}  \; \forall \, \mathbf{k}$, the scheme is unconditionally stable.


\subsection{Linearised Korteweg-de Vries equation}

The linearised Korteweg-de Vries (KDV) equation is given by
\begin{equation}
\frac{\partial u(\mathbf{x}, t)}{\partial t} = \frac{\partial^3 u(\mathbf{x}, t)}{\partial x^3} \enspace,
\end{equation}
where the Fourier representation of the spatial differential operator is $\lambda(\mathbf{k}) = -i k_x^3$. From Eq.\ \eqref{eq_kappa_t_kspace} the $k$-space correction term is
\begin{equation}
\hat{\kappa}_t(\mathbf{k}) = \frac{1 - e^{ -i k_x^3\Delta t} }{ i k_x^3 \Delta t}  \enspace.
\end{equation}
so the nonstandard PSTD scheme becomes
\begin{equation}
\frac{u^{n+1} - u^{n}}{\Delta t} = \mathcal{F}^{-1} \left\{ - \hat{\kappa}_t (\mathbf{k})i k_x^3 \, \mathcal{F}\left\{ u^n \right\} \right\} \enspace.
\end{equation}
As $k_x \in \mathbb{R}$, this means $\mathrm{Re}(\lambda(\mathbf{k})) = 0 \; \forall \, \mathbf{k}$ and the scheme is unconditionally stable.


\section{Application to PDEs with non-constant coefficients}

In the present work, only constant-coefficient PDEs in the form of Eq.\ \eqref{eq_general_pde} have been considered. However, with some constraints, the nonstandard PSTD method can also be used to derive approximate solutions to PDEs with spatially varying coefficients. In this case, a dispersion relation in the form of Eq.\ \eqref{eq_1st_order_general_dispersion_relation} or Eq.\ \eqref{eq_2nd_order_general_dispersion_relation} can no longer be formed. However, the $k$-space correction term derived for the corresponding constant-coefficient PDE can still be applied by choosing suitable reference values for the coefficients. As an example, consider the linearised wave equation with a non-constant sound speed
\begin{equation}
\frac{\partial^2 u(\mathbf{x}, t)}{\partial t^2} = c_0(\mathbf{x})^2 \nabla^2 u(\mathbf{x}, t) \enspace,
\end{equation}
where $c_0(\mathbf{x}) \in \mathbb{R}_{>0}$. Using the $k$-space correction derived for the constant-coefficient PDE given in Eq.\ \eqref{eq_kspace_correction_wave_equation}, the nonstandard  PSTD scheme becomes
\begin{equation}
\frac{u_m^{n+1} - 2 u_m^n + u_m^{n-1}}{\Delta t^2} = c_0(\mathbf{x})^2 \mathrm{F}^{-1} \{ - \hat{\kappa}_{tt}  k^2 \, \mathrm{F}\{ u_m^n \}   \} \enspace.
\label{eq_exact_PSTD_wave_equation}
\end{equation}
where 
\begin{equation}
\hat{\kappa}_{tt} =  \mathrm{sinc}^2(c_{\mathrm{ref}} k \Delta t/2)\enspace.
\end{equation}
Here $c_{\mathrm{ref}} \in \mathbb{R}_{>0}$ and typically $c_{\mathrm{ref}} \in c_0(\mathbf{x})$. This scheme was considered in \cite{Mast2001}. It is straightforward to write similar schemes for the other PDEs considered in Sec.\ \ref{sec_examples}. Note, in this case, the stability criteria depends on the values for the coefficients as well as the selected reference values.

The value of applying non-standard PSTD schemes to PDEs with non-constant coefficients can be explained by considering the role of the $k$-space correction term to be to reverse the numerical error introduced by the finite difference time step. Equations \eqref{eq_kappa_t_kspace} and \eqref{eq_kappa_tt_kspace} illustrate that this correction depends on both the size of the time step, as well as the values of the PDE coefficients. In the case of non-constant coefficients, the numerical error introduced by the finite difference time scheme will vary across the spatial domain depending on the local values of the coefficients. This means the correction will only be exact in regions where the PDE coefficients match the chosen reference values. However, if the variation of the coefficients across the domain is small, the $k$-space correction can still improve the stability and accuracy of the numerical scheme. This has previously been considered in the case of ultrasound wave propagation in soft biological tissue where the sound speed varies on the order of 5\% from the background values \cite{Treeby2012}. 


\section{Summary}

A class of exact PSTD schemes for solving time-dependent hyperbolic and parabolic PDEs is introduced. These schemes utilise the Fourier pseudospectral method for discretising spatial gradients and a nonstandard finite difference for time integration. The denominator function used in the construction of the nonstandard finite difference is transformed into the spatial frequency domain using the dispersion relation for the governing PDE. This allows the correction factor to be applied in the spatial frequency domain as part of the spatial gradient calculation. Using this approach, exact and unconditionally stable PSTD schemes can easily be obtained for a wide range of PDEs. An advantage of these schemes compared to nonstandard finite differences is that they apply to PDEs in any space dimension without further modification. These methods are likely to find applications in large-scale modelling problems, where counteracting the accumulation of phase errors becomes critical \cite{Treeby2012}.


\section{Acknowledgements}
This work was supported by the Engineering and Physical Sciences Research Council, U.K.

\end{document}